\documentclass{aa}
\usepackage{epsfig}
\def\ltsima{$\; \buildrel < \over \sim \;$}
\def\simlt{\lower.5ex\hbox{\ltsima}}
\def\gtsima{$\; \buildrel > \over \sim \;$}
\def\simgt{\lower.5ex\hbox{\gtsima}}
\def\srca{IRAS~14201+2956}
\def\srcb{IRAS~21582+1018}

\begin{document}

\title{ISO\thanks{An ESA project with instruments funded by ESA member
states.} observations of four active galaxies }
\author{M.Dennefeld\inst{1}\thanks{Visiting
astronomer, Observatoire de Haute-Provence (OHP), CNRS, France.}, 
T.Boller\inst{2}, D.Rigopoulou\inst{2}, H.W.W.Spoon\inst{3}}

\institute{
{Institut d'Astrophysique de Paris, 98bis Bd Arago, F-75014 Paris, France}
\and
{Max-Planck Institut for Extraterrestrial Physics, D-85740 Garching, Germany}
\and
{Kapteyn Astronomical Institute, 9700 AV Groningen, The Netherlands}
}
   
\offprints{M.Dennefeld}

\date{Received  ; accepted }

\authorrunning{M. Dennefeld et al.}

\markboth{M.Dennefeld et al.}{ ISO observations of four active  
galaxies }

\abstract{
We present ISO PHOT-S spectra of four galaxies known or suspected to host a central AGN.
Two of them are selected, among several others, from the
initial Iras/Rosat sample of Boller et al. (1992) because of their substantial 
X-ray emission, while no obvious  Seyfert features was present in their 
optical spectra:
\srca\ and \srcb\ . The latter, also known as Mrk~520, was bright enough to
also allow  SWS observations around selected neon lines, to establish its
excitation. While both PHOT-S spectra are characteristic of
starburst-dominated galaxies, the neon line ratios in \srcb\  indicate the
presence of a hard excitation source. Complementary optical spectra, both at
low and high spectral resolution, show only a weak, broad component around
H$_{\alpha}$, establishing the presence of a central AGN which may not be 
 detected in  standard, classification spectra. Both objects are now 
classified as Sey 1.9 galaxies.  These results show therefore that,
although IR observations were expected to be able to pierce through the dusty
central regions to reveal the presence of an active nucleus, the result may 
be ambiguous: 
the broad band IR energy distribution can still be 
  dominated by  starburts located in a wider circumnuclear region, and the 
AGN appear only in specific observations ( high-excitation lines in the IR, 
or high-resolution optical spectra).   
As a complement, two other galaxies from the same initial sample were also 
observed with PHOT-S:  the Narrow Line Seyfert galaxies (NLS1) 
Mrk~359 and Mrk~1388. 
NLS1s  appear in high proportion in the Rosat/Iras  
sample, and in soft X-ray samples in general, and  
   their Balmer line-widths 
are  sometimes  comparable to those of interacting, star-forming 
galaxies.   Their ISO
spectra  however do not reveal the typical, strong PAH features found in 
the starburst
galaxies and are more like those of  standard Seyferts. All  these  
observations   therefore indicate  that the key element is the
presence or absence of a circumnuclear starburst region which, if strong 
enough,
may completely hide   the presence of a
 central AGN in the IR spectral energy distribution . The dust  
obscuration  however needs to be patchy rather than complete to explain the 
detection of the high-excitation lines and Balmer wings in some cases.   
Only high-energy
observations can then establish the  strength of the central
 AGN and the amount of extinction with certainty.  
}

\maketitle

\section{Introduction}

The comparison between the ROSAT All Sky Survey (RASS)
and the IRAS Point Source
Catalog revealed many galaxies with X-ray luminosity in
the range $10^{42}$--$10^{43}$~erg~s$^{-1}$, which had not been
previously classified as Active Galactic Nuclei (AGN)
(Boller et al. 1992, Boller et al. 1998, hereafter B92 and B98). 
For many of them, this was simply due to the lack of 
optical spectroscopy, but for some others, the existing, 
low-dispersion spectra did not reveal  clear characteristics of a
Seyfert type nucleus. These cases opened the possibility that AGN may exist 
 in many galaxies without being detected by standard spectroscopy (Boller et 
al. 1993) and deserved further studies. As the selection process of
this sample included  detection by IRAS, it was also clear that obscuration
by dust was an important factor, as shown by the results 
obtained in various IRAS galaxies samples (e.g. Heckman et al. 1987; Veilleux 
et al. 1995). In particular, the controversy about the nature of the dominant
energy source, starburst or AGN, in the high-luminosity IRAS galaxies
(Sanders and Mirabel, 1996, and references therein) was illustrative of the
questions raised by IR-selected objects. 

%-----------------------------Figure 1--------------------------------
\begin{figure*}
\begin{center}
\epsfig{figure=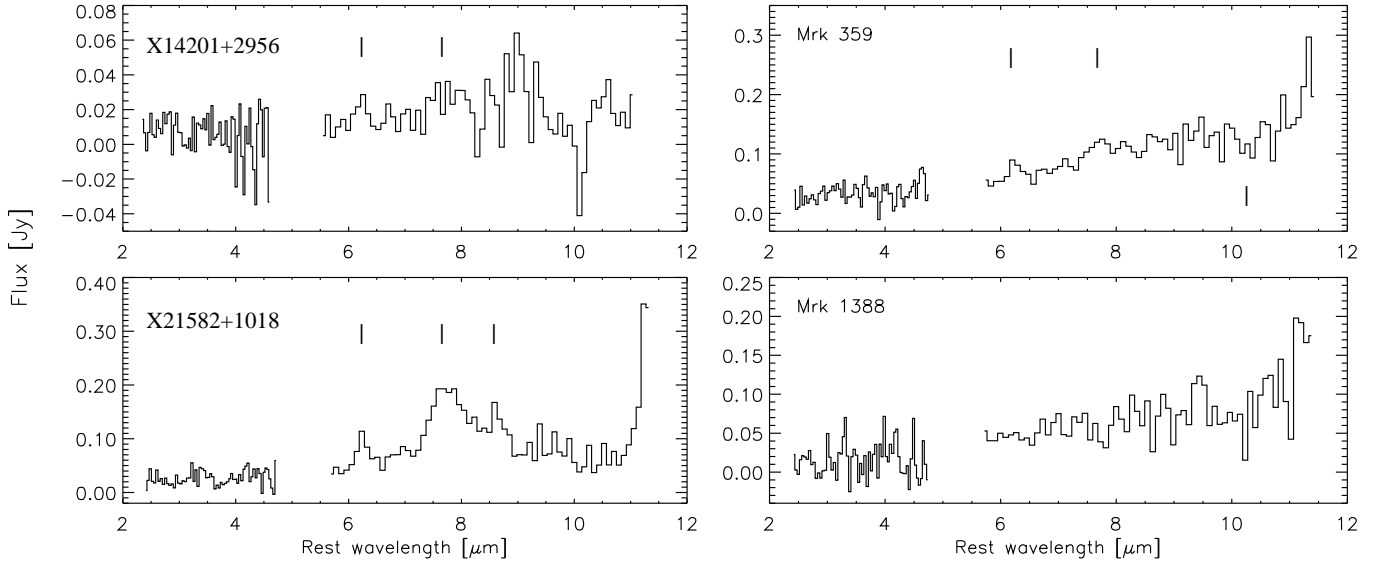,width=18.0cm,angle=00}
\end{center}
\caption{PHOT-S  spectra of the four galaxies. On the left, the two "starburst" 
galaxies, IRAS 14201+2956 and IRAS 21582+1018 (alias Mrk 520); on the right, 
the two NLS1 objects, Mrk 359 and Mrk 1388. Tickmarks indicate the detected 
lines. }
\label{fig1}
\end{figure*}
%-----------------------------Figure 1--------------------------------

A systematic optical spectroscopic follow-up of the B92 
sample was therefore undertaken, the results of which will be presented
elsewhere (Dennefeld et al., in preparation). 
Independant observations were also published by Moran et al. (1996). 
But at the same time, the
availability of the Infrared Space Observatory (ISO) of the European Space
Agency (Kessler et al. 1996) allowed  specific mid-infrared spectroscopy to
be envisaged for those objects with high X-ray luminosity  and no obvious AGN
signature in the visible, with the hope that the smaller extinction expected
at these IR wavelengths would allow a better determination of the
nature of the nuclear source. Observations of a large number of reference
objects, both starburst and Seyfert galaxies, in the ISO core-program, would 
provide the necessary reference for this identification: indeed, results by 
Genzel et al. (1998) show that both fine structure lines observed with the 
Short Wavelength Spectrometer (SWS) and polycyclic aromatic hydrocarbon (PAH)
features observed at lower spectral resolution with the ISOPHOT-S are 
excellent diagnostics to distinguish starburst and AGN energy sources. \\
The allocated observing time and the effective ISO sensitivity allowed us to 
observe only a few, bright objects, among the many initially selected. 
Two of them belong to the subclass of 
X-rays bright objects without previous Seyfert classification from the B92 
 sample: \srca\ and \srcb\ , the second one also being  known as 
Mrk~520 (Markarian and Lipovetskii, 1974). To those were added two other, 
bright, Narrow Line Seyfert  galaxies (NLS1): Mrk~359 (Markarian and 
Lipovetskii, 1971) and Mrk~1388 (Markarian et al., 1980). This class of  
objects  
was not well represented in the Genzel et al. (1998) core-program, but
appeared in large proportions in the B92 sample. The Balmer line-widths 
of NLS1 are much smaller than in standard Seyfert 1s, and are sometimes 
comparable to those of interacting, star-forming galaxies, 
 raising  the question of a 
possible link between the  peculiarities of the NLS1 and 
 the properties of dusty starbursts/AGNs.  \\
The basic parameters of these four objects are presented in Table~1. 
To facilitate the separation between the two types of objects, 
we will call the two NLS1s by their Markarian names, and refer to the two 
other objects by their IRAS name.  

%-----------------------------Table 1--------------------------------
\begin{table*}
\begin{footnotesize}
\begin{center}
\begin{tabular}{lcccccc} \hline 
Source & ${\rm \alpha_{2000}}$ & ${\rm \delta_{2000}}$ & ${V_\odot}$ & 
${L_{IR}}$ & ${L_X}$ (0.5--2~keV)  & IRAS  \\
& & & (km~s$^{-1}$) & ($10^{10}$~$L_\odot$)
& ($10^{42}$~ergs~s$^{-1}$) & ($25/60\mu$m flux ratio) \\ 
\ Mrk~359 & $01^{h}27^{m}32^{s}.3$ & $19^{o}$10$\arcmin$39$\arcsec$ & 5100  & 3.28 & 34.1 & 0.39  \\
\ Mrk~1388 & $14^{h}50^{m}37^{s}.8$ & $22^{o}$44$\arcmin$04$\arcsec$ & 6350 & 2.34 & 2.3  & 1.33  \\ 
\ \srca & $14^{h}22^{m}20^{s}.2$ & $29^{o}$42$\arcmin$55$\arcsec$ & 15700 & 15.8 & 13.7 & 0.21  \\
\ \srcb & $22^{h}00^{m}41^{s}.4$ & $10^{o}$33$\arcmin$08$\arcsec$ & 8200 & 23.1 & 5.7 & 0.12 \\ 
\hline 
\end{tabular}
\end{center}

\noindent

\caption{Basic properties of the four observed sources. \srcb~ is also known 
as Mrk 520. }
\label{tab1}
\end{footnotesize}
\end{table*}
%-----------------------------Table 1--------------------------------
 
In this paper, we report the results of the ISO observations of these four
targets, together with those of  complementary optical observations. 
 We will assume ${\rm H_0 = 50}$~km~s$^{-1}$~Mpc$^{-1}$, but the 
exact value has little influence on our conclusions.
Energies are quoted in the source rest frame . The ISO observations are
presented in Section~2, the other data  in Section~3 (X-rays, radio, optical),  
and the results are discussed in the last Section. 

\section{ISO observations and data reduction}

\subsection{The PHOT-S observations}

All four objects were observed in 1996-97 
with the ISOPHOT-S spectrophotometer (Lemke
et al. 1996) in rectangular chopped mode, with on-source 
integration times of 512 s. The total spectral range is covered
simultaneously by two linear
arrays, from 2.47 to 4.87$\mu$m  and from 5.84 to 11.62$\mu$m respectively,
with a common entrance aperture of 24" x 24". The ISOPHOT-S data were reduced
using the PIA version 8.1: the main steps of reduction can be found
in Rigopoulou et al. (1999). But it is worth mentioning here that, with this
 version, the short-wavelength range is also satisfactorily processed
and no additionnal flux correction factor is necessary. 
The resulting spectra are shown in Fig.~\ref{fig1} and some measured values
can be found in Table~2 (columns 4 to 8).

\subsection{SWS observations}

\begin{figure}
\begin{center}
\epsfig{figure=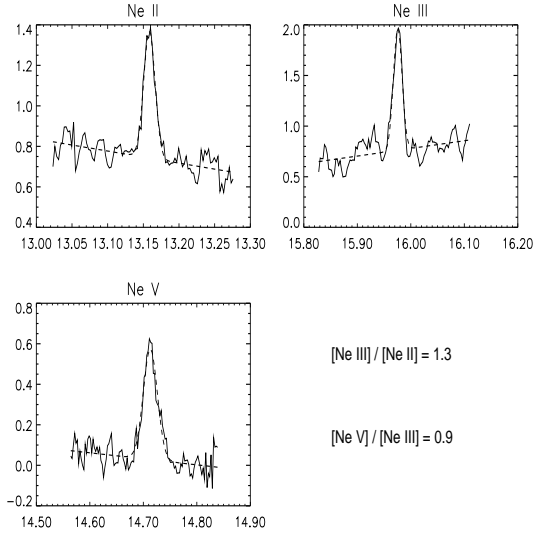,height=7.0cm,width=7.0cm,angle=-90}
\end{center}
\caption{Individual Neon lines detected in \srcb\ } 
\label{fig2}
\end{figure}

The brightest source, Mrk~520 = \srcb\ , was also  observed in late 1996 
with SWS,  in selected
wavelength ranges around the Neon lines ( [Ne~II] at 12.8, [Ne~V] at 14.3 and
[Ne~III] at 15.5$\mu$m rest wavelength respectively) in the high resolution 
SWS AOT6 mode. In this configuration, the entrance slit is 14" x 20", 
slightly  smaller than the aperture used in PHOT observations. 
The data were reduced using the SWS Interactive Analysis (IA) data reduction 
software package using standard ISO pipeline data routines and the 
corresponding calibration files (version 9.5). We have also used 
some more sophisticated software tools to improve dark current subtraction 
and flat fielding. The final spectra were re-binned to resolutions between
1200 and 1500.
The resulting lines together with Gaussian line fits (used to determine line 
fluxes) are shown in Fig.~\ref{fig2}.

%-----------------------------Table 2--------------------------------
\begin{table*}
\begin{footnotesize}
\begin{center}
\setlength{\tabcolsep}{1.0mm}
\begin{tabular}{lccccccc} \hline 
Source & ${\rm N_H}$ & ${\rm \Gamma}$ & {\rm 5.9 $\mu$m Cont.} & {\rm 7.7 $\mu$m Cont.} & {7.7 $\mu$m PAH Flux} & 
7.7 $\mu$m  & 6.2 $\mu$m PAH Flux \\
& $({\rm 10^{20}}$~$cm^{-2})$ & & (mJy)  & (mJy) & ($10^{-20}$~W~$cm^{-2}$) & L/C ratio & ($10^{-20}$~W~$cm^{-2}$) \\
\ Mrk~359 & $5.7(4.8){\dag}$ & $2.7 \pm 0.4$ & 53.1 $\pm 3.3$ & $81.2 \pm 7.4$ &  $12.2 \pm 1.7$ & $ 0.22 $ & $ 5.0 \pm 1.1$ \\ 
\ Mrk~1388 & $9.6(2.8)$ & $1.8 \pm 2.2$ & 43.0 $\pm 3.4$ & $53.7 \pm 7.6$ & $ \le 5.78$  &  & $ \le 3.5$ \\
\ IRAS 14201 & $1.7(1.3)$ & $1.7 \pm 0.8$ & 10.4 $\pm 2.1$ & $19.2 \pm 5.7$ & $ \le 3.74 $ &  & $ 2.3 \pm 0.9$ \\
\ IRAS 21582 & $20(5.4)$ & $3.0 \pm 2.5$ & 40.2 $\pm 2.9$ &  $55.8 \pm 6.7$ & $47.4 \pm 1.7$ & $ 1.42 $ & $ 12.7 \pm 1.2$ \\ \hline
\end{tabular}
\end{center}

\noindent
\hspace{2.5cm}$^{\dag}$Galactic value in parentheses
\caption{Best-fit parameters and results. The two first columns refer to the X-rays data, the last ones to the
ISO data.}
\label{tab2}
\end{footnotesize}
\end{table*}
%-----------------------------Table 2--------------------------------

\section{Additional data at other wavelengthes}

\subsection{X-rays data  }

The four objects were selected from  a cross-correlation (B92) between the 
Iras data 
and the Rosat All-Sky Survey; the latter is the main source for X-rays data. 
For each of the objects, a power law fit was done over the 0.1-2.4 keV spectral 
range, with free photon index $ \Gamma$  and hydrogen 
column density $ N_H$ . 
The results for $ N_H$ and  $ \Gamma$ are given in the first two 
columns of Table 2. The value for  the Galactic foreground absorbing column 
density (Dickey and Lockman, 1990, and references therein ) is given in 
parentheses after the value obtained from the best fit. 
 The absorption corrected flux is then used to 
derive the X-rays luminosity given in Table 1. \\
For Mrk 1388 and \srcb\, the 
spectral fit is poorly constrained. An alternative flux determination can 
then be done  by fitting a power law with fixed spectral index (2.3 
used here) and only the Galactic foreground absorbing column density. 
The resulting flux, 1.0 and 3.3 $10^{42}$~ergs~s$^{-1}$ respectively, is 
a factor of two lower than the result from the spectral fit, consistent with 
the lower absorbing column density used, and gives an estimate of the 
uncertainties. For \srcb\, the optical galaxy is offset from the centroid of 
the X-ray position but lies within the contours; 
   no other optical object is 
visible within these contours down to B=23 at least, and the galaxy is the 
only candidate, although the contribution from a neutron star cannot be 
formally excluded. 
\srca\ is in fact a pair of galaxies,
but the X-ray source is clearly related to one of the components only (the 
SW one),
 to which the ISO and optical spectra refer also.  No bridge is seen 
between those two galaxies on deep optical images obtained at OHP.  \\
No indication for variability is apparent from the Rosat data, except for 
\srca ~. This is also the only object for which other X-ray data are available: it 
has been observed in the Einstein extended medium-sensitivity survey (Gioia 
et al. 1990) with a flux of 
$(7.16 \pm 0.83)$~ $10^{-13}$~ergs~$cm^{-2}$~$s^{-1}$ 
in the 0.3-3.5 keV range. 
 The Rosat flux of  1.15 
$10^{-12}$~ergs~cm$^{-2}$~s$^{-1}$, in the  smaller spectral range of 
0.1-2.4 keV, is indeed higher, but this could be due to the soft-X excess 
only. We have used the power law fit done over the Rosat spectral range to 
evaluate the flux in the Einstein band (0.3-3.5 keV) and found a value 
of  5.9 $10^{-13}$~ergs~cm$^{-2}$~s$^{-1}$. The difference with the 
original Einstein observations appears to be marginally 
significant and could be  indicative of time variations also. \\
No hard X-ray data are  available for the moment for any of 
those objects, making difficult the analysis of the energy source from the 
X-ray point of view alone.

\subsection{Optical observations  }
Complementary  spectra were obtained over several observing runs at
Haute-Provence Observatory (OHP) with the 1.93m telescope, both at low and 
high
dispersion. The first spectra, at 7 $\AA$ per pixel, were obtained for \srca\
 (Fig.~3)
and \srcb\ in May 1992, and did not show any clear evidence of a Seyfert 
nature. In the spectrum of \srcb\, some line ratios ([NII]/H$_{\alpha}$ or 
[SII]/H$_{\alpha}$) show higher values than in standard HII regions, but  
not high enough to firmly establish the presence of an AGN (they could as well be 
produced by supernova-remnants). 
This surprising result  in view of the strong X-ray emission called for
additional efforts.  
Confirmation spectra at 7 $\AA$ per pixel 
were therefore obtained again,  as well
as spectra at a better resolution of 1.8 $\AA$ per pixel (in either the 
H$_{\alpha}$ or the H$_{\beta}$ spectral region). 
For \srcb\, this was done in September 1996:  Fig.~4 shows the low-dispersion 
spectrum, with better signal to noise than the 1992 one, while the 
high-dispersion one is presented in Fig.~5.  
In the mean time, Moran et al. (1996) published also a spectrum of this 
object, classifying it as a Sey 1.9~.
For \srca\,  a 
second low 
dispersion spectrum was secured in May 1995, but only the H$_{\alpha}$
region could be observed at higher resolution during this run. Both
the H$_{\alpha}$  and the H$_{\beta}$ region were re-observed during another
run in May 1998 and the H$_{\alpha}$ part is  presented in Fig.~6.  
The dashed lines in these figures represent the result of 
multi-component analysis, which will be discussed in  Section 4.2. 
 Low dispersion spectra were obtained for the known Seyferts Mrk 359 
(January 1996; Fig.~7) and   Mrk 1388 (June 1999; Fig.~8), to
check for eventual variability, but no conspicuous changes were noted. 
All spectra were reduced 
in the standard way and calibrated in flux with the help of standard stars
observed with the same setting during the same runs.  The relative fluxes are
estimated to be accurate within 10 $\%$ (except at the very 
blue end were the 
atmospheric extinction is not well calibrated).   Absolute
fluxes are not given systematically, as the observing conditions were not 
 perfect in each of the runs. 
 Small  variations of the total magnitude of an object 
from one run to the next can therefore not  be estimated  here.   
%-----------------------------Figure 2--------------------------------
%-----------------------------Figure 2--------------------------------

%-----------------------------Figure 3--------------------------------
\begin{figure}
\begin{center}
\epsfig{figure=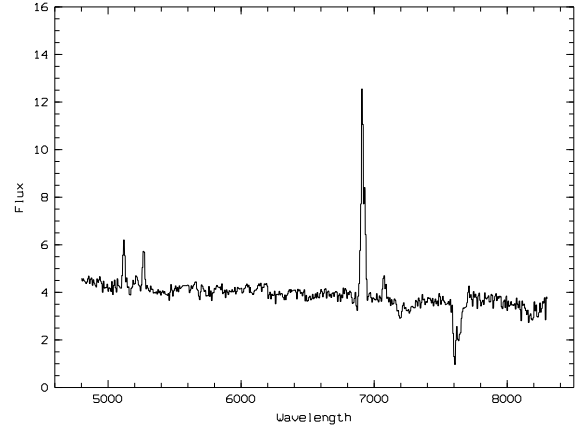,width=6.0cm,angle=-90}
\end{center}
\caption{Low dispersion spectrum of \srca\ obtained in May 92. Note the
absence of conspicuous Seyfert features.}
\label{fig3}
\end{figure}
%-----------------------------Figure 3--------------------------------

%-----------------------------Figure 4--------------------------------
\begin{figure}
\begin{center}
\epsfig{figure=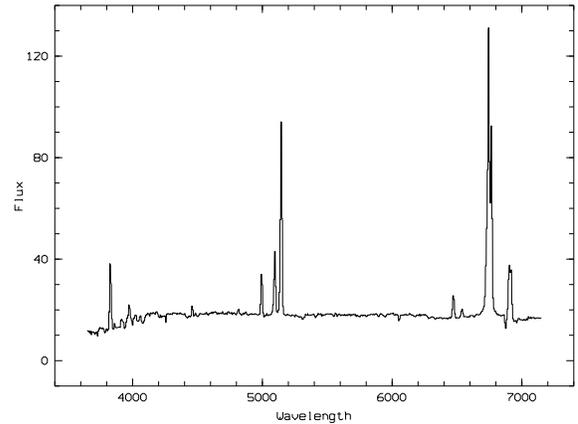,width=6.0cm,angle=-90}
\end{center}
\caption{Low dispersion spectrum of \srcb\ obtained in Sept. 96. No broad 
Aline features are apparent, but line ratios are close to the borderline 
between AGN and HII type galaxies. }
\label{fig4}
\end{figure}
%-----------------------------Figure 4--------------------------------

%-----------------------------Figure 5--------------------------------
\begin{figure}
\begin{center}
\epsfig{figure=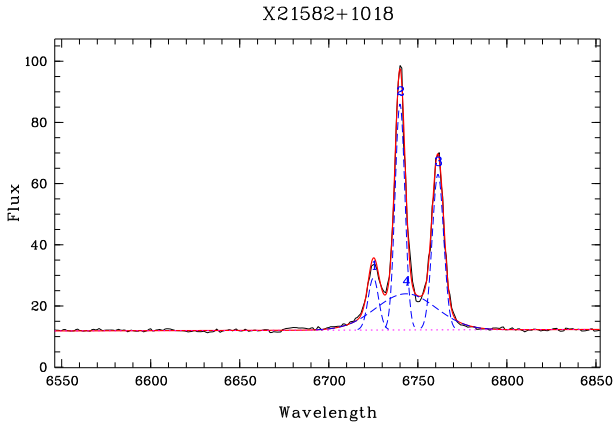,width=6.0cm,angle=-90}
\end{center}
\caption{ \srcb\ high  dispersion spectrum (Sept. 96) in the
 H$_{\alpha}$ region. 
The presence of a broad H$_{\alpha}$ component can only be 
established by multi-component fiting (dashed lines).}
\label{fig5}
\end{figure}
%-----------------------------Figure 5--------------------------------

%-----------------------------Figure 6--------------------------------
\begin{figure}
\begin{center}
\epsfig{figure=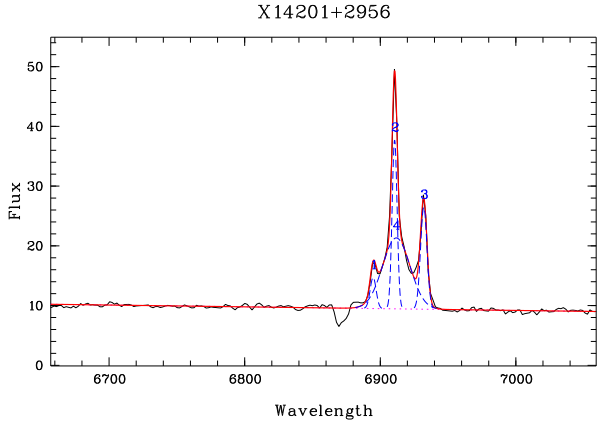,width=6.0cm,angle=-90}
\end{center}
\caption{ \srca\ high dispersion spectrum (May 1998)
in the H$_{\alpha}$ 
 region. The result of  multi-component fiting is shown with dashed-lines.}
\label{fig6}
\end{figure}
%-----------------------------Figure 6--------------------------------

%-----------------------------Figure 7--------------------------------
\begin{figure}
\begin{center}
\epsfig{figure=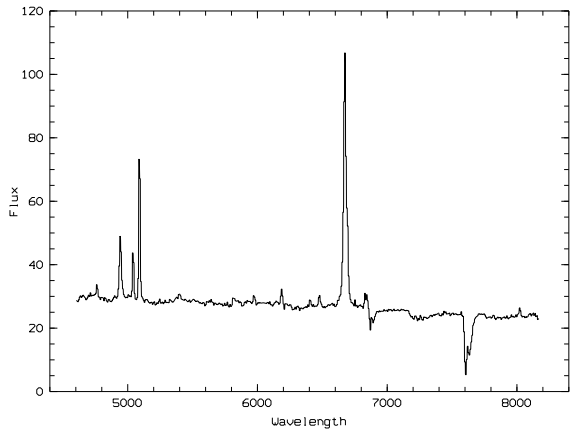,width=6.0cm,angle=-90}
\end{center}
\caption{ Low dispersion  spectrum of Mrk~359 (January 96). }
\label{fig7}
\end{figure}
%-----------------------------Figure 7--------------------------------

%-----------------------------Figure 8--------------------------------
\begin{figure}
\begin{center}
\epsfig{figure=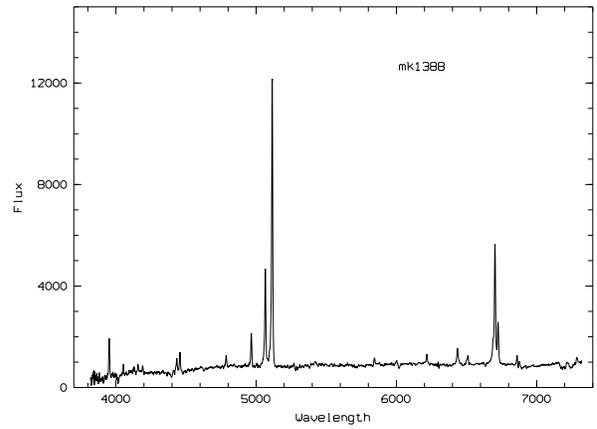,width=6.0cm,angle=-90}
\end{center}
\caption{ Low dispersion  spectrum of Mrk~1388 (June 99). }
\label{fig8}
\end{figure}
%-----------------------------Figure 8--------------------------------

\subsection{Radio data  }

We searched the publicly available  radio catalogues for counterparts to our 
objects.  \srcb\ was found in the NVSS survey (Condon et al. 1998) with a flux 
of 57.9 mJy at 1.4 GHz, and so was  the NLS1 Mrk359, with a flux of 4.8 mJy. 
 The other NLS1 in our list, Mrk 1388, 
was found in the FIRST catalogue as source J145037.8+224403 (the positional 
coincidence is better than 0"9 and is limited by the accuracy of the
 optical position) with an integrated flux of 9.93 mJy at 20 cm.    
 The source \srca\ is found to be associated with the source 
FIRST J142220.2+294255, with an 
integrated flux of 2.48 mJy at 20 cm.

\section{Discussion}

\subsection{The mid-IR properties  }

The basic properties of these galaxies can be best discussed with the PHOT-S
data, where all four objects have been observed. The spectrum with best
signal to noise, \srcb\, alias  Mrk~520, displays several PAH lines superposed 
on a 
relatively flat continuum. The strongest line is the 7.7$\mu$m, with a well
separated 8.6$\mu$m line on its red shoulder. The 6.2$\mu$m line is also well
detected. The sharp raise at the red extremity of the spectrum is clearly due
to the presence of the 11.3$\mu$m line, partly shifted out of the observed
range because of the recession velocity of this object. When comparing with
the various types of objects: starbursts, AGN's and ULIRG's, observed and
discussed by Lutz et al. (1998) (see for instance their Fig.~1), Mk~520
clearly appears like a typical Starburst galaxy in the mid-IR. Rigopoulou et 
al. (1999) also discussed a larger sample of ULIRG's, with special emphasis
on the possible effect of large extinction on the appearance of the mid-IR
spectrum. The main effect of extinction is to suppress the 8.6$\mu$m line and 
to depress the continuum redwards of it, as shown by a comparison between M82
and Arp~220 (their Fig.~6). No such suppression is seen here and the spectrum
of Mrk~520 clearly ressembles the one of M82 rather than Arp220. The 6.2/7.7
PAH flux ratio (0.27) is also consistent with a moderate
 extinction only. \\
 Although much noisier, the spectrum of \srca\ (the faintest of the four 
sources) resembles the one
of Mrk~520, with  detected 7.7 and 6.2$\mu$m PAH features and a flat
continuum. The 8.6$\mu$m line is not seen in the noise, while the 11.3$\mu$m
one is out of range due to the higher redshift. The silicate absorption is 
probably strong. This spectrum  is best
described as "Starburst" when comparing to the templates of Lutz et al.
(1998), but an "AGN" cannot be excluded.  \\
By contrast, the NLS1 Mrk~359 shows only  weak PAH features and a 
steadily raising continuum, typical of a power-law spectrum. 
This continuum is even detected in the usually noisy 2.4-5$\mu$m range.  
The 6.2$\mu$m PAH line is present with low contrast, as well as the 
7.7$\mu$ one.
 The 11.3$\mu$m line is  possibly seen at the edge of the spectrum and a weak 
silicate absorption may be present. This 
spectrum, with its higher slope and low contrast between line and continuum 
is  reminiscent of the AGN template in Lutz et al.. Although the presence
of the 6.2$\mu$m line indicates also a starburst contribution, this object 
is clearly AGN dominated. \\
 The last
spectrum, Mrk~1388, which has the lowest signal to noise ratio, does not show 
any
clear PAH line, but its   raising continuum 
indicates preferentially an AGN.  The slope of the continuum may be 
  affected by a
strong silicate absorption, as  indicated by the sharp rise at the red end. 
 The apparent "emission" around 9.4$\mu$m,  
which does not correspond to any known emission line, is probably instrumental 
in origin.   \\

We observed also  the Pf$\alpha$ line  in \srca ~ with SWS, in the hope to
reveal broad wings, signature of a Seyfert 1. We however failed to detect the 
line, and derived an 
 upper limit to its flux of $1.21 \times 10^{-20}   
W cm^{-2}$, with an uncertainty of about 20 percents. \\
Beyond the detection or not of the PAH features, the only other
mid-IR information  available for an identification of the nuclear 
source comes from the neon fine-structure lines, detected in Mrk~520, and 
displayed in Fig.~2. All three lines: [NeII] at 12.8$\mu$m, [NeIII] at 
15.5$\mu$m and [NeV] at 14.3$\mu$m are well detected with a good signal to
noise ratio. The important result  is that the [NeV]/[NeII] ratio is 
measured at
1.2, higher than in any other template observed by Genzel et al. (1998).
Although this ratio is not corrected for reddening (as no estimate of the
latter is available), its value will not change significantly for any
plausible reddening value (only 10$\%$ for a screen ${A_v}$ of 50)
and is  a clear sign of the presence of a hard UV radiation field,
i.e. an AGN. For Mrk~520, the PAH features and the fine structure lines
provide therefore apparently contradictory diagnostics.

\subsection{Comparison between IR and other wavelengths}

Putting aside for a moment the two NLS1 objects, the reason to observe the 
two other, "starburst", objects (\srca~ and \srcb ~) 
was  their high X-rays   
luminosity in the Rosat sample, rather uncommon for starburst objects.   
No known starburst galaxy is so X-ray luminous (Ptak et al. 1999). 
The mean relation between soft X-rays  and FIR luminosities  
derived empirically by David et al. (1992) for starburst and normal galaxies 
predicts an  X-ray luminosity one order of magnitude lower than observed for 
\srcb~ and two orders of magnitude lower for  \srca~, which is far beyond the 
scatter of this relation.  
The observed X-ray luminosities, $\sim 6$  and $\sim 14 
\times 10^{42}$~erg~s$^{-1}$ respectively, are, on the other hand, 
not uncommon among Seyfert galaxies (in our list, Mrk 1388 has in fact a 
lower X-ray luminosity than these two objects!). 

The low dispersion optical spectra do not show conspicuous signs of a Seyfert 
nature for \srca~ and \srcb~ (this was  the starting problem when
they were detected in the Rosat survey). In the course of this work, 
it appeared that \srca~ had been detected in the 
Einstein Medium Sensitivity Survey (Gioia et al. 1990), 
and quoted as an AGN without further precision. The spectroscopic follow-up 
by Stocke  et al. (1991) is  based on a blue spectrum only (without 
H$_{\alpha}$), and does not give a more precise classification:  as judged 
from our own spectra at similar resolution, this object was therefore among 
their 8 percent of objects where the classification was based on 
[OII]/[OIII] ratios only, 
and thus requiring additional observations.  No further spectrum was taken 
by Moran et al. (1996) in their Rosat follow-up, so it is not clear where  
 their S1 classification is coming from.  
Only the higher dispersion 
spectra presented here show unambiguously  the presence of a broad component 
around H$_{\alpha}$ (Fig.~4): 
  this object  can 
therefore be classified now as a Sey 1.9, as, indeed, no similar broad component 
is detected around H$_{\beta}$ (although the detection limit might be improved). 
 The excitation measured by the 
[OIII]/H$_{\beta}$ ratio is around 1, therefore a priori excluding a  
classical Seyfert 2 galaxy. \\

 For \srcb, no previous spectra were available and our first classification 
spectrum did not reveal broad components,  but an excitation  close to 
the border line between starburst and Seyfert galaxies.  A better spectrum 
obtained in 1996 (Fig.~4) confirms those line ratios, placing the object
 close to the border line but on the Seyfert side on the diagnostic 
diagrams presented by Veilleux et al. (1999) for Iras galaxies. The decisive   
ratio is the [OIII]5007$\AA$/H$_{\beta}$ one with a value of 5, while no 
broad H$_{\alpha}$ component is seen, therefore pointing towards a Seyfert 
2. The 
presence of the broad H$_{\alpha}$ component is only revealed in our 
high-dispersion spectrum  by a 
multi-component analysis: three narrow components with similar width 
(H$_{\alpha}$ and the two [NII] lines) and one H$_{\alpha}$ component with 
greater width were adjusted to reproduce the observed complex. The result of 
the fit is shown in Fig.~5 with dashed lines: the broad H$_{\alpha}$ 
component has an amplitude 7 times smaller than the narrow component and 
a FWHM of 1800 km/s. The observed width of the narrow components is 
310 km/s, while  the instrumental profile is 180 km/s FWHM.  
The total intensity of the broad H$_{\alpha}$ component 
is about equal to that of the narrow component. This object is thus 
also classified here as a Sey 1.9, in accordance with the classification 
proposed by Moran et al. (1996)  from a spectrum at $\sim 5$ $\AA$ 
resolution. 
 From the ratio of the narrow components of 
H$_{\alpha}$/H$_{\beta}$, we derive a reddening of $ A_{V} = 1.4$ magnitudes 
under the assumption of case B recombination.  

A similar multi-component analysis made for \srca~  (Fig.~5)
gives an intensity for the broad H$_{\alpha}$ component of 2.2 times the 
intensity of the narrow component, and a FWHM of only 1030 km/s (220 km/s 
for the narrow components). The 
H$_{\alpha}$/H$_{\beta}$ ratio gives here a reddening essentially zero, with 
however a significant uncertainty.   The optical 
classification of both objects as Sey 1.9 is thus now  in agreement 
with the detection of a strong soft X-rays component. \\ 
For the NLS1  objects, our spectra do not reveal any significant new feature 
compared to previous classification. Mrk~359 was observed in 
details by  Veilleux (1991), who noted its small line widthes and the 
apparent absence of reddening in the narrow line region.  
For Mrk~1388, the excitation is extremely high, a feature already noticed 
by Osterbrock (1985), and which is more 
appropriate for Seyfert 2 galaxies, than for  Seyfert 1s.  
No substantial reddening is indicated here for those objects either. 
 
While the X-ray and optical observations are now in agreement with an AGN 
classification for \srca~ and \srcb,  
 the contradiction of the broad IR features with this interpretation 
remains. The PAH features  detected by ISO-PHOT  characterise 
starburst galaxies. 
 The broad SED in the far IR, from 10 to 
100$\mu$m,   is also typical of 
starburst galaxies and does not satisfy the various criteria defined to 
select AGN in IRAS data (de Grijp et al. 1985; D\'esert \& Dennefeld 
1988). \\
On the contrary, both Mrk 359 and Mrk 1388, the NLS1 galaxies, 
 stand out with a mid-IR 
excess around 25$\mu$m, typical of warm dust heated by an AGN, the excess 
being particularly strong for Mrk 1388.  

If we use the radio data, to compute   the standard IR/radio parameter q  
discussed by Condon at al.(1995), we find  
a value of 2.52 for \srca~, again typical of starburst galaxies and showing 
that the AGN, if present, is not dominating 
the IR emission and/or that the object is radio-weak. 
 For Mrk 359, the q parameter has the value 2.51, which is not typical 
of a strong AGN, and thus points   probably towards a  
 radio-weak nature. 
 For Mrk 1388, on the 
contrary, this ratio 
is much lower (1.54), indicating the predominance of the AGN. The case 
of \srcb~ is intermediate, with q = 2.01, reflecting the complexity of 
this source and the mix of starburst and AGN. \\
Finally, the 
observed infrared luminosities for the two Sey 1.9s, 
\srca~ and \srcb~, are very high  
if we assume that the objects have an  
 ${\rm L_{IR}/L_X}$ ratio typical for AGN or quasars with
${\rm L_X < 10^{45}}$~erg~s$^{-1}$ ($\simeq$4.5; Elvis et al. 1994):  
extrapolating the Rosat fluxes and correcting for absorption provides 
a 0.5--10~keV luminosity of 4 and $ 2 \times 10^{44}$~erg~s$^{-1}$, which 
translates into an ${\rm L_{IR}/L_X}$ ratio of 16 and 45 respectively. 
 The 
bulk of the IR emission has  therefore to be attributed  to another source 
than the AGN. 
The fact that \srca~ is located close to another galaxy (included in the 
Iras lobe) could mean that this second object is also contributing to the 
far IR emission, but we can rule out a significant contribution from it 
because its 
optical spectrum is not the one of a typical Iras starburst galaxy (no 
emission lines present).  

The discrepancy between X-ray and IR signatures is therefore best resolved 
if the two emissions  
have a different origin. The X-rays come from a central AGN, whose 
contribution to the IR emission is negligible. 
The bulk of the IR radiation is   produced by an
intense star formation episode, occurring on spatial scales of the
order of or larger than the NLR, and not directly linked to the AGN. 
The limited angular resolution of our ground-based 
spectra (1"2 per pixel, with an average seeing of 2") does not allow us to 
disantangle spectroscopically the AGN from the starburst region: we can 
only say that this star forming region must be contained within a "central" 
region of about 3 kpc diameter for  \srca ~ and  2 kpc for \srcb. This is 
completely consistent with what is known for other, starbursting, IRAS 
galaxies. 
  However the fact that 
the neon line ratios  clearly indicate an AGN in \srcb ~ means that the 
dusty starburst is not fully obscuring the view to the central region. 
The other question  is then to understand why  
the optical spectrum hardly reveals the broadline region, while the 
X-rays are coming out:   
the measured optical extinction of $ A_{V}$~ $\sim 1.4$~ is 
largely unsufficient to hide it. 
 The column density derived from the X-ray spectral analysis provides 
only 
$A_{V} = 1$, barely consistent with the value derived from the optical: 
there is 
therefore no room in the X-rays for additional extinction towards the center. 
The measured optical extinction probably refers then to an outer region 
only (starburst/NLR), possibly linked to a dusty, warm absorber, as known in 
other cases (e.g. Reynolds et al., 1997). This is another argument for  a 
distinct origin of the IR and X-ray emission. 
For \srca~, both the optical and the X-ray indicate an absorption close to 
zero, but the broad H$_{\alpha}$ component is also better detected. \\
We can also use an isotropic indicator of the AGN's intrinsic brightness,  
like the [OIII] luminosity, corrected for extinction, and compare it to 
the 1--10~keV luminosity 
(derived by extrapolation from the Rosat data) to get an estimate of the 
absorption affecting the X-rays flux (Maiolino et al. (1998); Bassani 
et al. (1999). Using our 1996 spectrum  for 
\srcb~ (obtained in photometric conditions), we derive  a 
${\rm L_X/L_{OIII}}$ ratio of 1.3 only, on the very low side of the bulk 
of Seyfert 1 galaxies (Maiolino et al. 1998, their Fig.~6), thus  
leaving room for at least  
a  moderate absorption, although a compton-thick source is difficult 
to justify with the available data (as a comparison, the same  ratio 
derived for Mrk 1388 
is about 10, a value typical for  Sey 1 galaxies). For \srca~, this ratio is 
much higher 
($\sim 200$): even if it  is affected by a large uncertainty (a factor of 2) 
on the [OIII] luminosity, this object is probably not  heavily absorbed. 
In both cases however,  
one has to remember that the X-rays fluxes used in the calculation are 
extrapolated from soft 
data, and could therefore be overevaluated due to  the possible presence 
of a strong soft X-rays excess. 

We could therefore propose  that these two objects are better described as 
weak AGNs rather than highly obscured objects. 
 From their optical properties, they refer mainly 
to the Sey 1.9 or 1.8 classes, where the reason for weak broad lines is 
still not understood.  As they have a large soft X-ray component, 
 they could 
  be closer than suspected to NLS1s, where the extinction is small 
also.   It 
will therefore be important to get hard X-ray observations for all these 
objects, to determine the relative importance of the soft X-ray component 
(excess or not), to check the existence of a dusty, warm absorber and 
to assess the 
real strength of the central AGN and the obscuration in front of it.
Confirming the possible variability is another element to clarify their 
nature. They may in the end be local examples of the absorbed type-2 objects 
searched for at much higher redshift.

\section{Conclusion}

A combination of optical, IR and X-ray observations has helped to 
clarify the apparent discrepancies in the properties of the two galaxies 
\srca~  and \srcb~. The high soft X-ray emission detected by Rosat, which 
is difficult to explain by a starburst, is linked to a central AGN whose 
presence is detected in the optical only by high resolution  spectroscopy 
revealing a weak, broad component in  H$_{\alpha}$. The  emission and 
spectral energy distribution in the IR, on the 
other hand, is clearly dominated, even in the ISO mid-IR range, 
 by a circumnuclear starburst revealed by the PAH features. The discrepancy 
between the X-ray and IR properties is therefore best explained by a 
different origin of the main emissions. In the case of  \srcb\, however, 
high resolution spectroscopic observations with ISO have also revealed 
high-excitation lines clearly associated with the AGN and its NLR, but not 
dominating the energy balance in the IR. 
The extinction measured from the X-ray or the optical is insufficient to 
explain the lack of strong, broad emission lines and  pertains 
essentially to the starburst/NLR only. 
 \srcb\, and to a lesser extent \srca\, are therefore best 
described by a central, 
perhaps weak, AGN, with a BLR partly obscured by a structured absorber, and an   
absorbed NLR, mixed with a region of intense star formation of perhaps larger 
extension.  
These examples  show that the detection of AGN spectral line  
characteristics in the optical or IR are not enough to establish the main 
source of energy in the IR and that mixed cases may be frequent. Many of 
them are probably found in the B92 and B98  samples, as the 
selection there  was made both by X-ray and IR emission. The only established 
common property between those objects and the NLS1 galaxies  also found 
in large number in these samples (of which two 
examples, Mrk~359 and Mrk~1388, were studied here)
seems to be the strong soft X-ray emission. Whether or not 
both type of objects have more common properties, and are 
for instance both  characterized by  a weak central AGN remains 
to be established: only hard X-ray observations will be able to 
  measure the strength 
of the central engine and determine the surrounding absorption.  

\begin{acknowledgements}
  
 This research
has made use of the LEDA Extragalactic Database, which is operated
at Observatoire de Lyon (France), and of data
obtained through the High Energy Astrophysics Science Archive Research
Center Online Service, provided by the NASA/Goddard Space Flight Center.
We thank the anonymous referee for helpful comments.

\end{acknowledgements}

\end{document}